\documentclass[12pt]{article} 
 
\usepackage{epsfig} 
 
\newcommand{\xbj}{x}

{\end{list}} 
\newcounter{enumct}

\newcommand{\captive}[1]{\rule{5mm}{0mm}%
\begin{minipage}{150mm}\caption[small]{#1}\end{minipage}}

\begin{document} 
\input feynman 
\bigphotons 
\sloppy 
 
\pagestyle{empty} 
\begin{center} 
{\LARGE\bf Resolved virtual photons in the small $x$ domain }\\[4mm] 
{\Large H.~Jung, L.~J\"onsson, H.~K\"uster} \\[3mm] 
{\it Department of Physics,}\\[1mm] 
{\it Lund University, 221 00 Lund, Sweden }\\[1mm] 
{\it E-mail: jung@mail.desy.de, leif@quark.lu.se}\\[20mm] 
{\bf Abstract}\\[1mm]  
\begin{minipage}[t]{140mm} 
It has been found that recent results on forward jet production  
from deep inelastic scattering can neither be reproduced  
by models which are based on leading order $\alpha_s$ QCD matrix  
elements and parton showers 
nor by next-to-leading order calculations. 
The measurement of forward jet cross sections has been 
suggested as a promising probe of  
new small $\xbj$ parton dynamics 
and  
the question is 
whether these data provide an indication of this. 
The same question arises for 
other experimental data in deep inelastic scattering at small  
$\xbj$ which  
can not be described by conventional models for deep inelastic scattering.   
In this paper the influence of resolved  
photon processes has been investigated and it has been studied to  
what extent the inclusion of such processes in addition to normal 
deep inelastic scattering 
leads to agreement with data. 
It is shown that two DGLAP  
evolution chains from the hard scattering process towards the proton 
and the photon, respectively, 
 are sufficient to describe effects, observed 
in the HERA data, which have been attributed to BFKL dynamics. 
\end{minipage}\\[5mm] 
 
\rule{160mm}{0.4mm} 
 
\end{center} 
 
\section{Introduction} 
Experimental data from deep inelastic scattering (DIS) 
in a kinematic region where new parton dynamics is expected 
to become noticeable i.e. at small values of the scaled proton  
momentum, $x$, are not described by models based on interactions 
with pointlike photons. 
In a previous paper \cite{JJK_resgamma} we have demonstrated that the addition of 
interactions through resolved photons offers a possible explanation 
to the observed discrepancies and leads to good agreement with all 
available data.

This paper is devoted to a more detailed discussion of the resolved 
photon concept and comparisons with data on 
forward jet production in DIS since  
the forward jet cross section has been 
advocated as a particularly sensitive measure of  
small $\xbj$ parton dynamics 
 \cite{Mueller_fjets1,Mueller_fjets2}. 
Analytic calculations based on  
the BFKL equation in the   
leading logarithmic approximation (LLA) 
 are in 
fair agreement with data.  
However, recent calculations of the BFKL kernel in 
the next-to-leading logarithmic approximation (NLLA) \cite{BFKL_NLO} 
have given surprisingly large corrections, and it remains to be shown whether 
the data can still be reasonably described. 
\par 
Monte Carlo generators  
based on direct, point-like  
photon interactions (DIR model), 
 calculated from leading order (order $\alpha_s$)  
QCD matrix elements, and leading log parton showers  
based on the DGLAP evolution  do not 
take  
any new parton dynamics in the small $\xbj$ region 
into account and are therefore not expected 
to fit the experimental data.  
Recent results from the H1~\cite{H1_fjets_data} 
and ZEUS~\cite{ZEUS_fjets_datab} 
experiments on forward jet production  
exhibit significant deviations from the predictions of such 
models.  
Also next-to-leading order calculations (NLO i.e. order $\alpha_s^2$)  
predict too small a cross section compared to data.  
\par 
The study of forward jet production with contributions from 
direct as well as resolved photon processes has been performed using 
the RAPGAP~2.06~\cite{RAPGAP,RAPGAP206} Monte Carlo event  
generator.  
 
\section{Resolved Photons in DIS} 
In electron-proton scattering the internal structure of the proton as  
well as of the exchanged photon can be resolved provided the scale of  
the hard subprocess is larger than the inverse radius of the proton,  
$1/R^2_p \sim \Lambda_{QCD}^2$, and the photon, $1/R^2_{\gamma} \sim Q^2$, 
respectively. Resolved photon processes play an important role in  
photo-production of high $p_T$ jets, where $Q^2 \approx 0$,  
but they can also give considerable contributions to DIS  
processes \cite{H1_incl_jets,Chyla_res_gamma}  
if the scale $\mu^2$ of the hard  
subprocess is larger than $Q^2$, the inverse size of the photon. 
\par 
In the following we give a brief description of the model for resolved virtual 
photons used in the Monte Carlo generator RAPGAP. 
Given the fractional momentum transfer of the incoming electron to the  
exchanged photon, the Equivalent Photon Approximation provides the  
flux of virtual transversely polarized photons  
\cite[and references therein]{RAPGAP206,RAPGAP}. 
The contribution from longitudinally polarized photons has been neglected. 
The partonic structure of the virtual photon is defined by  
parameterizations of the parton densities, 
$x_{\gamma} f_{\gamma}(x_{\gamma},\mu^2,Q^2)$, which   
depend on the two scales $\mu^2$ and $Q^2$ 
\cite{GRS,Sasgam,Drees_Godbole}.  
The following hard subprocesses are considered (RES model):  
$ gg \rightarrow q \bar{q}$, 
$ g g \rightarrow gg$,  
$ q g \rightarrow q g $,  
$ q \bar{q} \rightarrow g g $,  
$ q \bar{q} \rightarrow q \bar{q}$, 
$ q q \rightarrow q q $.  
Parton showers on both the proton and the photon side are included. 
The generic diagram for the process $ q_{\gamma} g_{p} \rightarrow q g $ 
including parton showers is shown in Fig.~\ref{resgam1}. 
 
\begin{figure}[ht] 
\begin{center} 
\begin{picture}(30000,25000) 
\drawline\fermion[\NE\REG](5000,25000)[5000] 
\drawline\fermion[\E\REG](0,25000)[5000] 
\drawline\photon[\S\REG](5000,25000)[3] 
\global\advance\pmidx by -5000 
\put(\pmidx,\pmidy) {$y,Q^2 \to$ } 
\drawline\fermion[\E\REG](\photonbackx,\photonbacky)[1500] 
\global\advance\pmidy by 400 
\global\advance\fermionbackx by +1000 
\global\advance\pbackx by 500 
\global\advance\pbacky by -3200 
\global\advance\Yfive by + 3800 
\drawline\fermion[\S\REG](\photonbackx,\photonbacky)[1000] 
\global\advance\pmidx by -4000 
\drawline\gluon[\E\REG](\fermionbackx,\fermionbacky)[2] 
\global\Xseven = \pbackx 
\global\Yseven = \pbacky 
\global\advance\Xseven by + 500 
\global\advance\Yseven by - 300 
\put(\Xseven,\Yseven){$q_{T\;i}$} 
\global\Xone = \pbackx 
\global\Yone = \pbacky 
\global\advance\Xone by + 4500 
\global\advance\Yone by - 750 
\put(\Xone,\Yone){{\Huge  
   $\searrow$} \hspace{0.5cm} $x_{\gamma}f(x_{\gamma},\mu^2,Q^2)$  
        \hspace{0.2cm} $\gamma$ - DGLAP} 
\drawline\fermion[\S\REG](\fermionbackx,\fermionbacky)[1500] 
 
\drawline\gluon[\E\REG](\fermionbackx,\fermionbacky)[3] 
\global\Xeight = \pbackx 
\global\Yeight = \pbacky 
\global\advance\Xeight by + 500 
\global\advance\Yeight by - 300 
\put(\Xeight,\Yeight){$q_{T\;i+1}$} 
 
\drawline\fermion[\S\REG](\fermionbackx,\fermionbacky)[2500] 
\global\advance\pmidx by -5000 
\drawline\fermion[\E\REG](\fermionbackx,\fermionbacky)[4500] 
\global\Xeight = \pbackx 
\global\Yeight = \pbacky 
\global\advance\Xeight by + 500 
\global\advance\Yeight by - 300 
\put(\Xeight,\Yeight){$p_{T}$} 
\global\advance\pmidy by 400 
\global\advance\pbackx by 500 
\global\advance\pbacky by -1500 
\global\Xsix = \pbackx 
\global\Ysix = \pbacky 
\global\advance\Ysix by + 2800 
\drawline\gluon[\S\REG](\fermionfrontx,\fermionfronty)[2] 
\global\Xtwo = \pmidx 
\global\Ytwo = \pmidy 
\global\advance\Ytwo by - 500 
\global\advance\Xtwo by - 5500 
\put(\Xtwo,\Ytwo){{$\mu^2$, $\hat{t} \to $}} 
\global\advance\Xtwo by + 6500 
 
\global\advance\Xtwo by + 6500 
\put(\Xtwo,\Ytwo){{\Huge $\} $}} 
\global\Xthree = \Xtwo 
\global\advance\Xthree by + 3000 
\global\advance\Ytwo by + 1000 
\put(\Xthree,\Ytwo){ $qg \to qg$} 
\global\Ythree = \Ytwo 
\global\advance\Ythree by - 1500 
\put(\Xthree,\Ythree){ hard scattering } 
\drawline\gluon[\E\REG](\gluonbackx,\gluonbacky)[4] 
\global\Xeight = \pbackx 
\global\Yeight = \pbacky 
\global\advance\Xeight by + 500 
\global\advance\Yeight by - 300 
\put(\Xeight,\Yeight){$p_{T}$} 
\global\advance\pmidy by 400 
\global\advance\pbacky by +400 
\global\advance\pbacky by -1500 
\drawline\gluon[\S\REG](\gluonfrontx,\gluonfronty)[3] 
\global\advance\pmidx by -3000 
\drawline\gluon[\E\REG](\gluonbackx,\gluonbacky)[3] 
\global\Xeight = \pbackx 
\global\Yeight= \pbacky 
\global\advance\Xeight by + 500 
\global\advance\Yeight by - 300 
\put(\Xeight,\Yeight){$q_{T\;j+2}$} 
\drawline\gluon[\S\REG](\gluonfrontx,\gluonfronty)[2] 
\drawline\gluon[\E\REG](\gluonbackx,\gluonbacky)[2] 
\global\Xeight = \pbackx 
\global\Yeight= \pbacky 
\global\advance\Xeight by + 500 
\global\advance\Yeight by - 300 
\put(\Xeight,\Yeight){$q_{T\;j+1}$} 
\drawline\gluon[\S\REG](\gluonfrontx,\gluonfronty)[2] 
\drawline\gluon[\E\REG](\gluonbackx,\gluonbacky)[1] 
\global\Xeight = \pbackx 
\global\Yeight= \pbacky 
\global\advance\Xeight by + 500 
\global\advance\Yeight by - 300 
\put(\Xeight,\Yeight){$q_{T\;j}$} 
\global\Xfour = \pbackx 
\global\Yfour = \pbacky 
\global\advance\Xfour by + 2500 
\global\advance\Yfour by + 2000 
\put(\Xone,\Yfour){{\Huge  
   $\nearrow$}  \hspace{0.5cm}$x_{p}f(x_{p},\mu^2)$  
                \hspace{0.2cm} $p$ - DGLAP} 
 
\drawline\gluon[\SW\REG](\gluonfrontx,\gluonfronty)[2] 
 
\global\advance\gluonbackx by -1500 
 
\multiput(\gluonbackx,\gluonbacky)(0,-1000){3}{\line(1,0){9000}} 
\global\advance\gluonbacky by -1000 
\global\advance\gluonbackx by -500 
\put(\gluonbackx,\gluonbacky){\oval(1000,3000)} 
\global\advance\gluonbackx by +2000 
\global\advance\gluonbacky by +1000 
\global\advance\gluonbackx by -2000 
\global\advance\gluonbacky by -1000 
\global\advance\gluonbackx by -500 
\drawline\fermion[\W\REG](\gluonbackx,\gluonbacky)[2000] 
\global\advance\fermionbacky by -3000 
\global\advance\fermionbackx by 4000 
\global\advance\fermionbacky by 3000 
\global\advance\fermionbackx by -4000 
\global\advance\pmidy by 500 
\put(\pbackx,\pmidy){$p$} 
\end{picture} 
\end{center} 
\captive{Deep inelastic scattering with a resolved virtual photon and 
the $q_{\gamma} g_p \to q g $ partonic subprocess. 
\label{resgam1} } 
\end{figure} 
 
Since the photon structure function depends on the scale, $\mu^2$, of the  
hard scattering process, the cross section of resolved photon processes  
will consequently also depend on the choice of this scale.  
It has to be carefully considered in which range of $\mu^2/Q^2_0$ 
the photon-parton cross section can be factorised into a parton-parton 
cross section convoluted with the parton density of the photon. 
The parton density of the photon is evolved from a starting scale $Q^2_0$ 
 to the scale $\mu^2$, the virtuality  
at the hard subprocess, giving a resummation to all orders. 
 
\subsection{Parton Distribution Functions} 
Due to factorization of the cross sections the parton densities of both the 
virtual photon and the proton enter into the calculations. The proton  
structure function, $F_2$, has been measured to high accuracy and therefore the 
various parameterizations only give marginal differences in the 
measurable kinematic region. 
Two parameterizations of the parton distribution in the proton,  
GRV 94 HO (DIS) and CTEQ4D have been considered,  
which both give good agreement with 
the proton structure function 
data \cite{H1_F2,ZEUS_F2}. It was found that 
the produced results were identical within the percent level, 
when keeping $\Lambda_{QCD}$ fixed. In the following we use only GRV HO.  
\par 
The photon can interact via its 
partons either in a bound vector meson state or as decoupled partons if  
the $p_T$ of the partons is high enough.  
The splitting 
$\gamma \to q \bar{q}$  is called  
the anomalous component of the photon. 
The structure function of virtual photons 
has only recently been measured but by far not to the same precision as the 
proton structure function. However, it turns out that data are in good 
agreement with the parameterization of Schuler and Sj\"ostrand (SaS)  
~\cite{Sasgam}.  
The SaS parameterization  
offers a choice of $Q_0^2$ values at which the anomalous part becomes effective.  
We have studied  
these choices resulting in different magnitudes of the 
parton densities, and consequently of the cross sections.  
For the SaS  
parameterization  we have used  $Q_0^2$ as given by eq.(12) of ref.~\cite{Sasgam} 
 ($IP2 = 2$). This choice is also suitable for a description of other hadronic
 final state properties (not considered in this paper),
  like energy flow, forward particle spectra and jet
 cross sections.
\par 
In a previous paper \cite{JJK_resgamma} we have shown that the 
ansatz of Drees and Godbole \cite{Drees_Godbole} gave very similar results  
to those obtained from  
the SaS parameterization why we restrict ourselves to the SaS   
parton distributions in this study. 
\par 
The hadronic contribution to the virtual photon structure function 
decreases rapidly with increasing $Q^2$, which means that the main contribution at 
large $Q^2$ comes from the anomalous piece in the photon splitting. This 
is completely calculable in pQCD and leads to an expected agreement between  
the parameterization of Gl\"uck - Reya - Stratman~\cite{GRS} and 
that of Schuler - Sj\"ostrand~\cite{Sasgam} but 
also with the simple ansatz of Drees - Godbole \cite{Drees_Godbole}. 
However differences 
exist in the way the hadronic part of the structure function is matched to the 
pointlike part, which just reflects the theoretical uncertainty. 
 
\subsection{Choice of Scale} 
In leading order $\alpha_s$ processes 
the renormalization scale $\mu_R$   
and factorization scale $\mu_F$ are not well defined 
which allows a number of reasonable choices. 
There are essentially two competing effects: 
a large scale suppresses $\alpha_s(\mu^2)$ but 
gives, on the other hand, an increased 
parton density, $xf(x,\mu^2)$, for a fixed small $x$ value. 
The net effect depends on the details of the interaction  
and on the parton density parameterization. 
\par 
In previous papers~\cite{Jung_resgamma,JJK_resgamma} 
we have tried different scales like $\mu^2= 4 \cdot p_T^2$ and $\mu^2 = Q^2 + p_T^2$, 
and found that these choices gave similar results.   
\par 
However, in resolved virtual photon processes the choice of 
the scale $\mu^2$, at which the photon is probed, is severely 
restricted \cite{gosta_torbjorn}. 
In a partonic process $ a + b \to c + d$, where $a, b, c, d$ 
denote four-vectors and where parton $a$ has the virtuality $Q^2$,  
the transverse momentum $p_T^2$  
of parton $c$ is given in the small angle limit ($-\hat{t} \ll \hat{s}$) by: 
$p_T^2 = \hat{s}(-\hat{t})/(\hat{s} + Q^2) $, with $\hat{s}$ and 
$\hat{t}$ being the usual 
Mandelstam variables. In a $t$ channel process the  
virtuality is given by $\mu^2 = -\hat{t}$. 
Thus we have:\footnote{We are grateful to T. Sj\"ostrand for pointing out this simple 
explanation} 
\begin{equation} 
\mu^2 = -\hat{t} = p_T^2 + Q^2 \cdot \frac{p_T^2}{\hat{s}} < Q^2 + p_T^2 
\label{kin-constraint} 
\end{equation} 
 From eq.(\ref{kin-constraint}) we see, that the scale $\mu^2$ is always larger than 
the transverse momentum squared of the hard partons and less than $Q^2 + p_T^2$. 
In the following we will use $\mu^2 = Q^2 + p_T^2$ as scale for both resolved  
virtual photon processes and for direct photon processes. 
This choice of scale provides a smooth transition 
from the kinematic region of normal DIS into the range where resolved  
photons start contributing and to the photo-production region.  
The same scale has also been used in 
NLO calculations including resolved photons in deep inelastic scattering  
\cite{Kramer_Poetter_dijets}.  
\par 
A basic test that the scale is reasonable is that the 
parton shower evolution scheme should not be able to produce partons with  
transverse momenta larger than those produced 
by the matrix element for the hard scattering process. 
 
\section{Forward Jets} 
HERA has extended the available $\xbj$ region down to values  
below $10^{-4}$ where new parton dynamics  
might show up. 
Based on calculations in the LLA of the BFKL kernel, the cross section for 
DIS events 
at low $\xbj$ and large $Q^2$ with a high $p^2_T$ jet in the  
proton direction (a forward jet) \cite{Mueller_fjets1,Mueller_fjets2} is 
expected to rise more rapidly with decreasing $\xbj$ than expected 
from DGLAP based calculations. 
New results from the H1~\cite{H1_fjets_data} and  
ZEUS~\cite{ZEUS_fjets_datab} 
experiments have recently been  
presented. 
The data can be described neither by conventional DIR Monte Carlo models 
nor by a NLO calculation, while 
comparisons to analytic calculations of the 
LLA BFKL 
mechanism has proven reasonable agreement. 
\par 
It should be kept in mind that both the  
NLO calculations and the BFKL based 
calculations are performed on the parton level whereas the  
data are on the level of hadrons.  
\par 
\begin{figure}[htb]
\begin{center}
\epsfig{figure=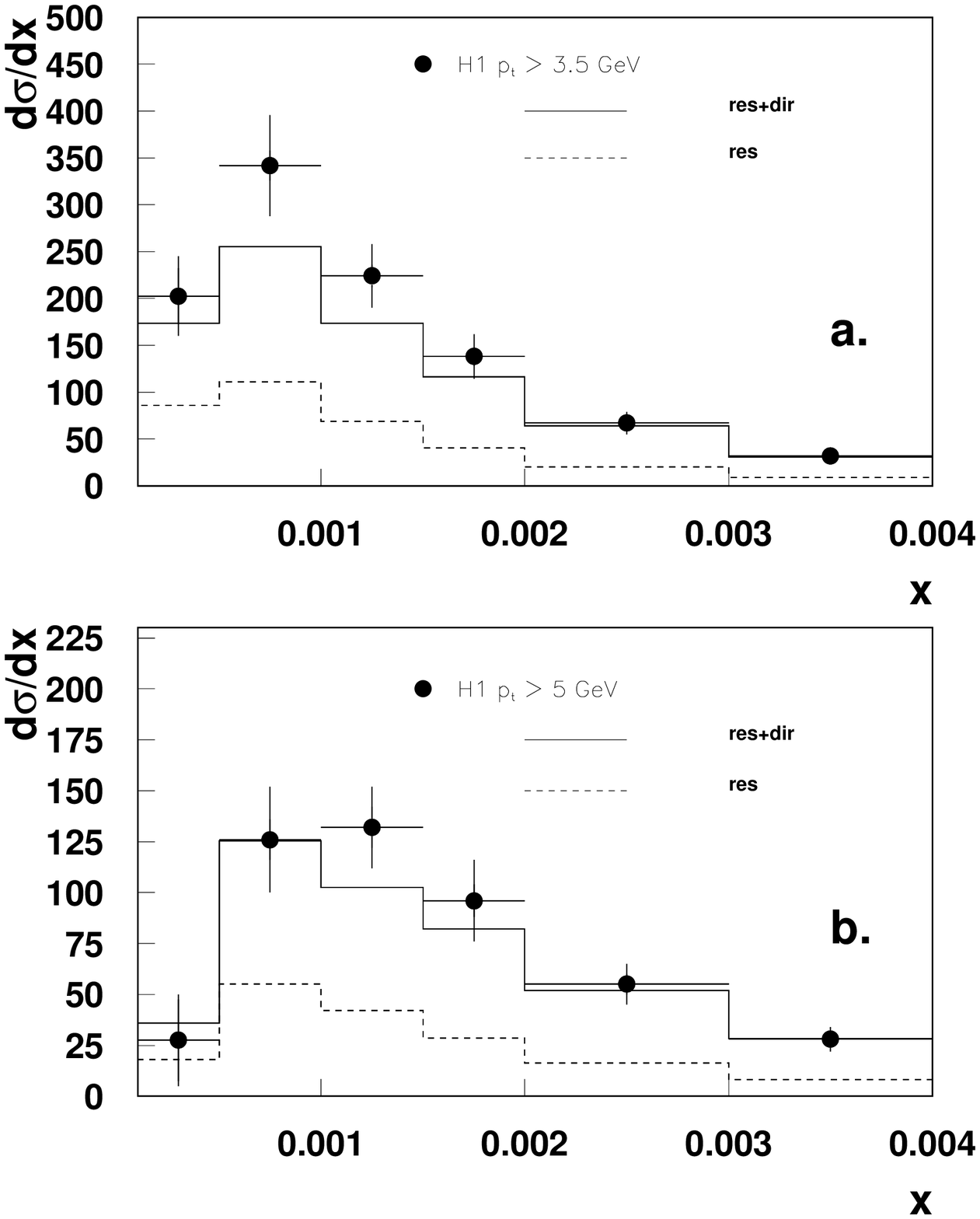,
width=15cm,height=15cm}
\end{center}
\captive{
The forward jet cross section as a function of $\xbj$ 
for $p_{T\; jet} > 3.5$ GeV ($a.$) and $p_{T\; jet} > 5$ GeV ($b.$).
Also shown are
the RAPGAP predictions 
for the sum 
of direct and resolved processes (solid line) as well as the
resolved photon contribution alone (dashed line).
\label{fjet_res+dir}}
\end{figure}

In Fig.~\ref{fjet_res+dir} the forward jet  
cross section as measured by the    
H1 collaboration  \cite{H1_fjets_data} is   
compared to the prediction of the RAPGAP Monte Carlo generator for both,  
resolved photon process alone (labeled RES) and for 
the sum of direct and resolved processes (labeled DIR+RES). 
The calculation is performed with a   
scale $\mu^2=Q^2+p_T^2$ used in the parton densities and $\alpha_s$  
for both direct and resolved 
photon processes. The measurement is well described by the Monte Carlo, 
including direct and resolved photon processes.  
The forward jet cross section as measured by the ZEUS experiment 
\cite{ZEUS_fjets_datab} 
can be equally well described by this Monte Carlo program using the  same  
structure functions and parameter setting.  
 
\begin{figure}[htb] 
\begin{center} 
\epsfig{figure=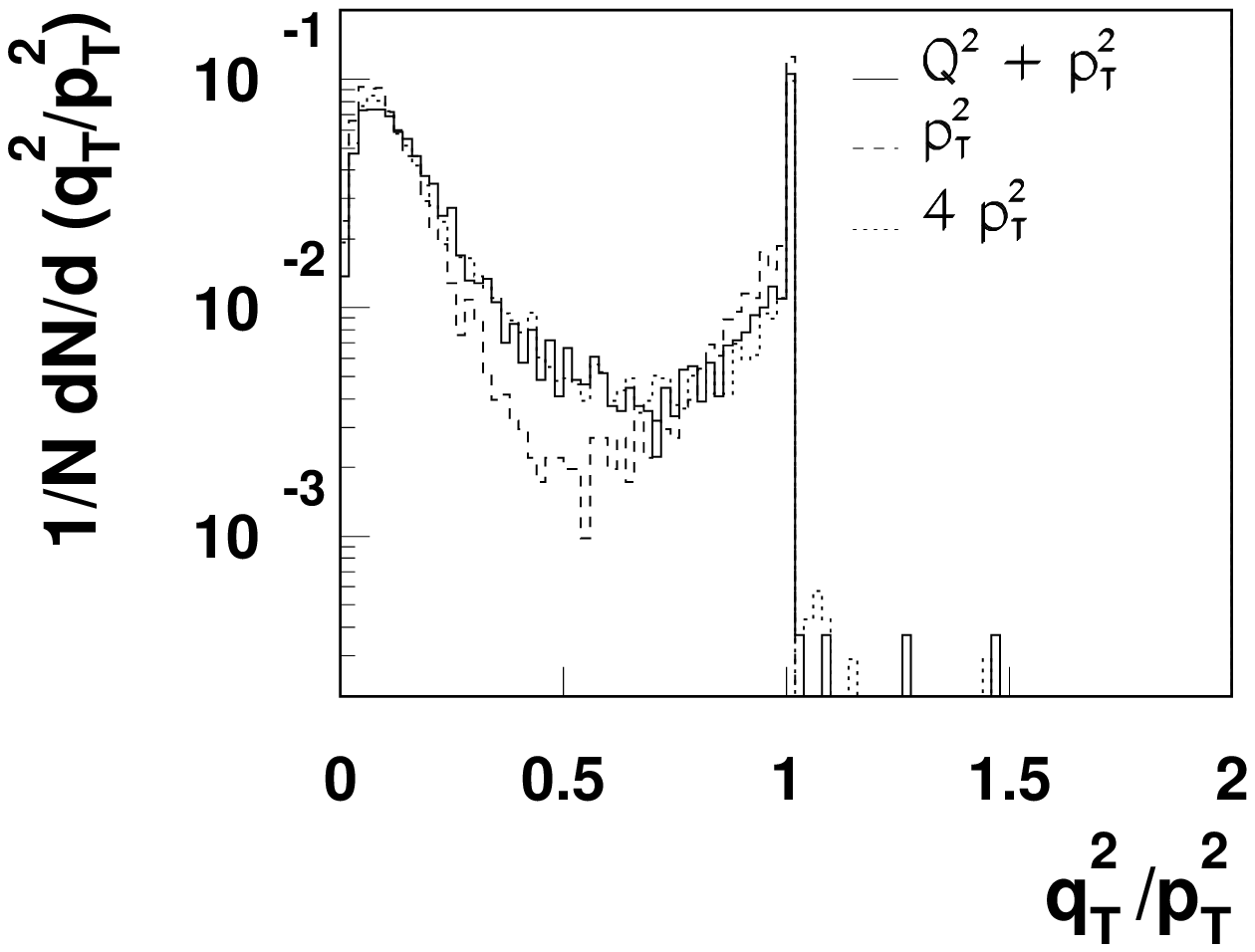, 
width=13cm,height=13cm} 
\end{center} 
\captive{The ratio $R=q^2_T/p^2_T$ 
 of the transverse momenta $q^2_T$ of partons from the initial state 
cascade to the transverse momentum $p^2_T$  
of the partons from the hard scattering 
process. the solid line corresponds to the scale $\mu^2=Q^2+p_T^2$, the dotted 
line to 
$\mu^2=4 \cdot P_T^2$ and the dashed line to $\mu^2=p_T^2$. The distribution is 
normalized to the total number of entries. Please note the logarithmic scale on 
the $y$ - axis. 
\label{fjet_ptgluon}} 
\end{figure} 
For the forward jet analysis it is important to check that 
the reconstructed jets really stem from the hard scattering and not  
from the initial state parton cascade. 
In Fig.~\ref{fjet_ptgluon} the ratio of the transverse momentum of any 
initial state parton $q_T^2$ to the transverse momentum $p_T^2$  
of the hard scattering process is shown 
in the $\gamma ^* p$ CMS for events which satisfy the forward jet analysis 
criteria. The solid line corresponds to $\mu^2=Q^2 + p_T^2$, the dotted line to 
$\mu^2=4 \cdot p_T^2$ and the dashed line to $\mu^2=p_T^2$. One can see, that 
essentially all partons coming from the initial state cascade have  
transverse momenta smaller than the partons of the hard scattering $p_T$,  
which is expected in a DGLAP type 
evolution, where the transverse momenta are ordered in $q_T$ towards the hard 
scattering process. Thus we conclude that the scale $\mu^2=Q^2 + p_T^2$  
which has been used for this study fullfils the  
requirements of a DGLAP type initial state cascade. 
\begin{figure}[htb] 
\begin{center} 
\epsfig{figure=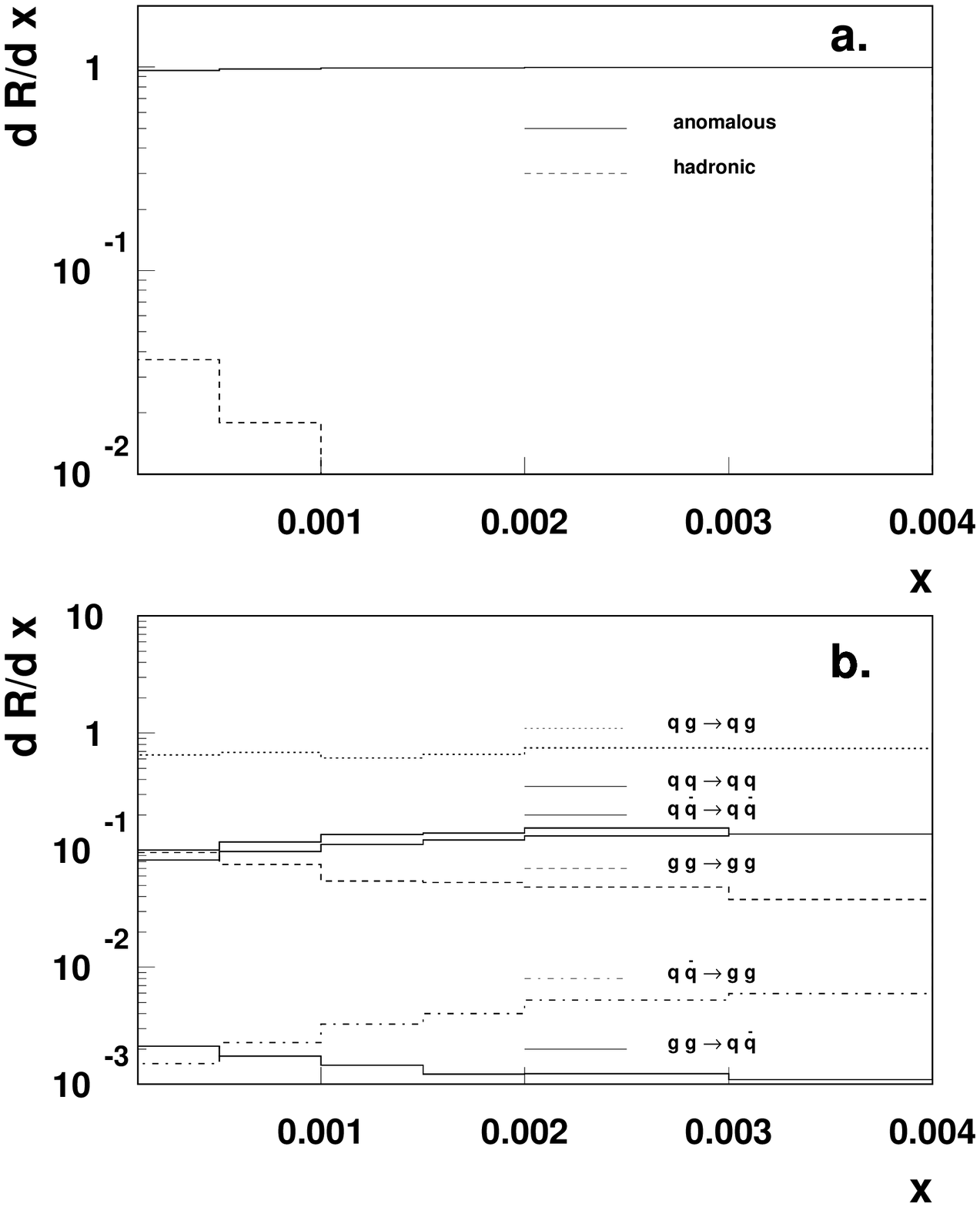, 
width=13cm,height=15cm} 
\end{center} 
\captive{Different contributions  to the total cross section of 
resolved virtual photons within the cuts of the 
forward jet analysis \protect\cite{H1_fjets_data}. 
In $a.$ is shown 
the ratio $R= \frac{\sigma_i}{\sigma_{res.\;tot}}$, i.e.   the
 pointlike part (solid line) and the  
hadronic part (dashed line), respectively,
of the resolved virtual photon cross section divided by the
 the total resolved 
photon cross section as a function of $x$. 
In  
$b.$ is shown the ratio $R=\frac{\sigma_i}{\sigma_{res.\;tot}}$ 
as a function of $x$
for different subprocesses $i$: $qg \to qg $ (dotted line),
$qq \to qq$ and $q \bar{q} \to q \bar{q}$ (solid line),
$gg \to gg$  (dashed line), $q \bar{q} \to gg$  (dashed-dotted line) and
$gg \to q \bar{q}$ (solid line).
\label{fjet_res+dir_process}} 
\end{figure} 
\par 
In Fig.~\ref{fjet_res+dir_process} the different contributions  
of the total  resolved photon cross section are shown separately  
within the cuts of the forward jet analysis. 
From Fig.~\ref{fjet_res+dir_process}$a$ it is observed that the hadronic  
part of the virtual photon structure function as expected gives a 
negligible contribution to the measured cross section, since  
it dies off rapidly with increasing $Q^2$. 
Fig.~\ref{fjet_res+dir_process}$b$ shows that the subprocess 
$q_{\gamma}g_p \to qg$  
contributes the most to the resolved photon cross section in the  
forward jet region ($\sim 60 \%$)  
and that  the subprocesses
$q q  \to q q $, $q \bar{q}  \to q \bar{q} $ and $g g \to g g$
each give a  contribution of the order of $10 \%$. 
\par 
A small fraction of the DIS events, fulfilling the selection criteria 
for forward jets, actually contains two identified jets.   
Analytic calculations (in LLA) \cite{BFKL_dijets} 
have been performed in the same kinematic 
region and with the same jet selection as defined for the one-jet sample. 
The predicted ratio varies from $3 \%$ to $6 \%$ as   
$\xbj$ increases from $0.5 \cdot 10^{-3}$ to $\xbj = 3 \cdot 10^{-3}$. 
Our previously reported prediction from the RAPGAP generator ~\cite{JJK_resgamma} 
including both direct and resolved photon processes was that about 1 \% 
of the total forward jet sample contains two forward jets.   
This is about a factor of 3 lower than the prediction from the  
BFKL calculations but 
a large part of this discrepancy could be due to hadronization effects 
which would reduce the prediction of the parton level BFKL calculation. 
\par 
Recently this ratio has been measured by the H1 experiment \cite{H1_fjets_data} 
to be 1 \%, in excellent agreement with the prediction of RAPGAP. 
This gives further confidence in the basic concept of resolved photons 
even at large $Q^2$. 
\par
\begin{figure}[htb]
\begin{center}
\epsfig{figure=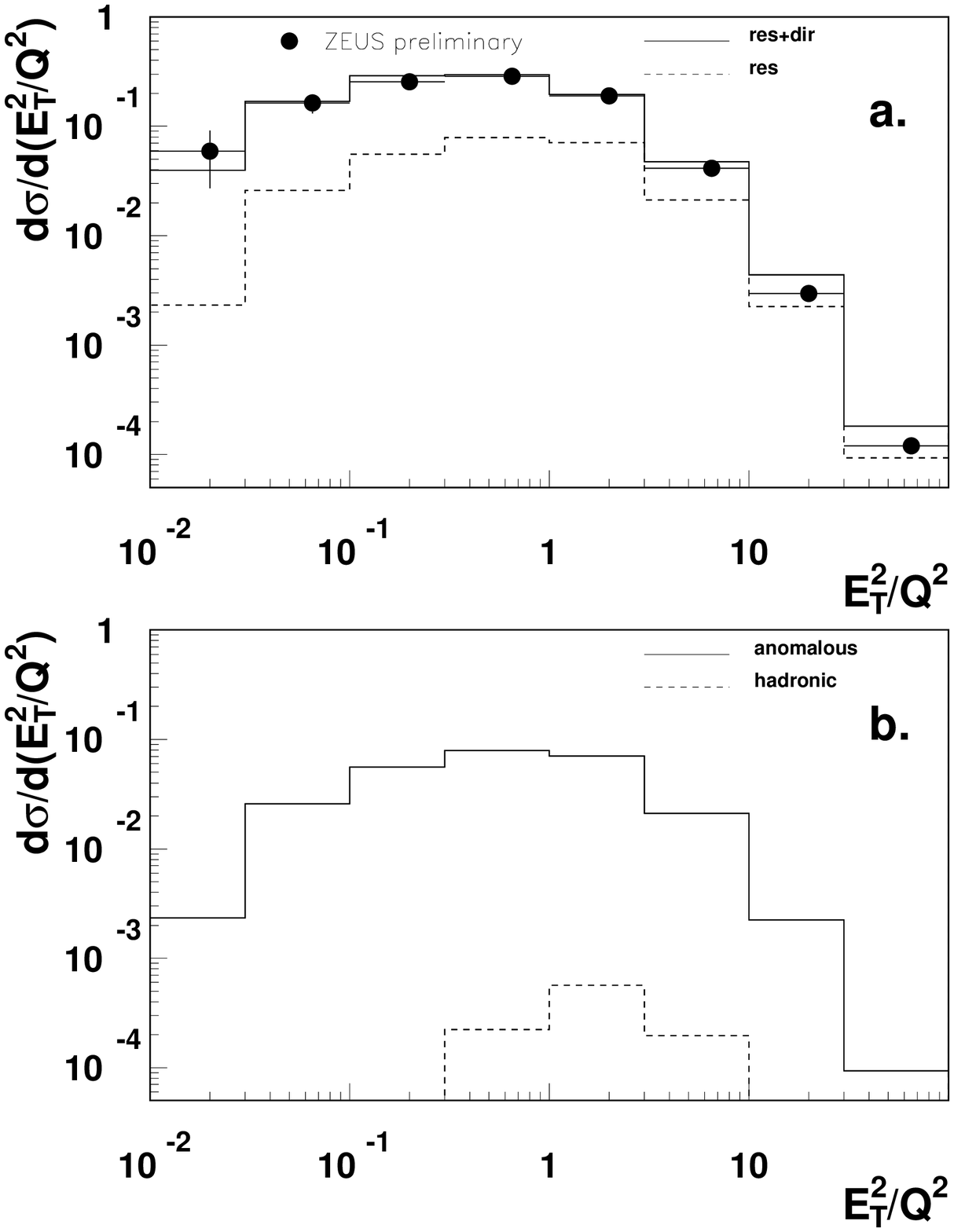,
width=15cm,height=15cm}
\end{center}
\captive{The cross section of forward jets as a function of $E_T^2/Q^2$.
In $a.$ is shown the measurement of 
ZEUS~\protect\cite{ZEUS_fjets_pt2/q2} and  the prediction
of RAPGAP. 
The solid line shows the sum of direct and resolved virtual photon
contributions,
whereas the dashed line shows the resolved  photon contribution alone.
In $b.$ is shown the part of 
the cross section
coming from the anomalous component (solid line)
and the one coming from the hadronic component of the virtual photon
separately. 
\label{fjet_pt2/q2}}
\end{figure}

 
The ZEUS collaboration has presented a measurement of  
the forward jet cross section as a function of 
$p_T^2/Q^2$~\cite{ZEUS_fjets_pt2/q2}  
in the kinematic region $Q^2>10$ GeV$^2$, $y>0.1$, $E_{e'} > 10$ GeV, 
$\eta_{jet}<2.6$, $x_{jet} >0.036$, $E_{T\;jet}> 5$ GeV, $p_{z\;breit} > 0$ GeV, 
$2.5 \cdot 10^{-4} < x < 8 \cdot 10^{-2}$ but  
without implementing the DGLAP suppression cut  
$0.5 < E_T^2/Q^2 < 2$. The results are compared to the  
predictions from different Monte Carlo programs, and the conclusion  
is that only the RAPGAP DIS generator including interactions through 
resolved virtual photons can describe the data over the full 
range in $E_T^2/Q^2$. In Fig.~\ref{fjet_pt2/q2}$a$ the 
ZEUS results are shown together with the prediction of RAPGAP and in  
Fig.~\ref{fjet_pt2/q2}$b$ the contributions coming from the hadronic  
and the pointlike part of the virtual photon structure function  
are presented separately.  

\section{Summary and Discussion} 
 
Recent experimental data on forward jet production show deviations from 
traditional LO Monte Carlo models assuming   
directly interacting point-like photons. 
It is tempting to assume that the observed effects could be explained by 
BFKL dynamics.  
\par 
In the present study we have shown that the addition of resolved  
photon processes to the direct interactions in DIS 
leads to good agreement with the data. 
This agreement does not depend on any specific choice of scale or  
tuning of any other parameters in the RAPGAP generator. The best  
evidence of the universality of this approach is that, with the  
same parameter setting, it is possible to describe a wide range of 
other data like the transverse energy flow~\cite{H1_energyflow}, 
transverse momentum spectra of single particles~\cite{H1_ptspectra_data}, 
the (2+1) jet rate~\cite{H1_2+1jets_data}  
and single inclusive jet cross sections~\cite{H1_incl_jets}, as we have shown 
in \cite{JJK_resgamma}. 
\par   
We have observed that the dominant contributions to the resolved photon 
processes come from order $\alpha_s^2$ diagrams with the hard  
subprocess $q_{\gamma} g_p \to q g $ (see Fig.~\ref{resgam1}). 
 Since the partons which form 
the photon remnant per definition have smaller $p_T$ than the  
partons involved in the hard scattering, a situation with non $q_T$ 
ordering is created. 
\par 
In the LO DIR model the ladder of gluon emissions is  
governed by DGLAP dynamics giving a strong ordering of  
$q_t$ for emissions between the photon and the proton vertex. 
The models describing resolved photon processes and BFKL dynamics  
are similar in the sense that both lead to a breaking of this 
ordering in $q_t$. 
The BFKL picture, however, allows for complete dis-ordering in 
$q_t$, while in the resolved photon case the DGLAP ladder is split into 
two shorter ladders, 
one from the hard subsystem to the proton vertex, and one to the photon 
vertex, each of them ordered in $q_t$ (see Fig.~\ref{resgam1}). 
 Only if the ladders are long enough to produce  
additional hard radiation it might be possible to separate resolved  
photon processes from processes governed by BFKL dynamics. 
Thus the resolved photon approach may be a ``sufficiently good'' 
approximation to an exact BFKL calculation and the two approaches may prove  
indistinguishable within the range of $\xbj$ accessible at HERA.  
\par 
It should be emphasized again that a NLO calculation  
assuming point-like virtual 
photons contains a significant part of what is 
attributed to the resolved structure of the virtual photon in the  
RES model~\cite{Kramer_Poetter_dijets}.

\section{Acknowledgments} 
 
It is a pleasure to thank G. Ingelman and A. Edin for discussions 
about the concept of resolved photons. We have 
also profited from a continuous dialogue with B. Andersson, G. Gustafson 
and T. Sj\"ostrand. We want to thank G. Kramer and B. P\"otter for many 
discussions on the relation between 
 resolved photons in DIS and its relation to NLO 
calculations.  
We also want to thank M. W\"usthoff and J. Bartels 
 for discussions on BFKL and the resolved photons in DIS.

\end{document}